\documentclass[12pt,preprint]{aastex}
\def\kms{km~s$^{-1}$}
\def\ms{m~s$^{-1}$}
\def\mjup{M$_{\rm Jup}$}
\def\mearth{M$_{\earth}$}
\def\msun{M$_{\odot}$}
\def\msini{$M_P\sin i~$}
\def\vsini{$V_{rot}\sin i~$}
\def\rphk{$\log{R^\prime_{HK}}$}
\def\fe{[Fe/H]}
\def\starA{GJ\,317}
\def\pA{692.9}
\def\peA{4}
\def\pyearsA{1.897}
\def\pyearseA{0.01}
\def\tpA{2451639}
\def\tpeA{30}
\def\eA{0.193}
\def\eeA{0.06}
\def\omA{344}
\def\omeA{10}
\def\kA{100}

\def\trendA{7.6}
\def\trendeA{1}
\def\msiniA{1.2}
\def\arelA{0.95}

\def\chisA{1.42}
\def\nobsA{18}
\def\mstarA{0.24}
\def\bvA{1.52}
\def\vmagA{12.0}
\def\mvA{12.2}

\def\feA{-0.23}
\def\parallaxA{109.}
\def\parallaxeA{20.}
\def\dA{9.17}

\def\kmass{0.26}
\def\vmass{0.22}

\def\kA{71.0}
\def\pBB{673.4}
\def\pBC{2700}
\def\msiniBB{1.3}
\def\msiniBC{0.83}
\def\chisq{$\sqrt{\chi_{\nu}^2}$}
\def\chio{1.63}
\def\chitr{1.42}
\def\chitwo{1.11}
\def\rmso{16.8}
\def\rmstr{12.5}
\def\rmstwo{6.32}

\def\fraclow{1.8}
\def\fraclowe{1.0}
\def\fracmed{4.2}
\def\fracmede{0.7}
\def\frachi{8.9}
\def\frachie{2.9}
\begin{document}
\title{A New Planet Around an M Dwarf: 
Revealing a Correlation Between Exoplanets and Stellar Mass$^1$}

\author{John Asher Johnson\altaffilmark{2},
R. Paul Butler\altaffilmark{3}, 
Geoffrey W. Marcy\altaffilmark{2},
Debra A. Fischer\altaffilmark{4},
Steven S. Vogt\altaffilmark{5},
Jason T. Wright\altaffilmark{2},
Kathryn M. G. Peek\altaffilmark{2}
}

\email{johnjohn@astro.berkeley.edu}

\altaffiltext{1}{ Based on observations obtained at the
W.M. Keck Observatory, which is operated jointly by the
University of California and the California Institute of
Technology. Keck time has been granted by both NASA and
the University of California.}

\altaffiltext{2}{Department of Astronomy, University of California,
Berkeley, CA USA  94720}

\altaffiltext{3}{Department of Terrestrial Magnetism, Carnegie Institution
of Washington, 5241 Broad Branch Road NW, Washington D.C. USA 20015-1305}

\altaffiltext{4}{Department of Physics and Astronomy,
San Francisco State University, San Francisco, CA, USA 94132}

\altaffiltext{5}{UCO/Lick Observatory, University of California
at Santa Cruz, Santa Cruz CA USA 95064}

\begin{abstract}
We report precise Doppler measurements of \starA\ (M3.5V) that reveal
the presence of a planet with a minimum mass \msini~$ =
$~\msiniA~\mjup\ in an eccentric, \pA~day orbit. \starA\ is only the
third M dwarf with a Doppler--detected Jovian planet. The residuals
to a single--Keplerian fit show evidence of a possible second orbital
companion. The inclusion of an additional Jupiter--mass
planet ($P \approx 2700$~days, \msini$= \msiniBC$~\mjup) decreases \chisq\
from \chisA\ to \chitwo, and 
reduces the rms from \rmstr~\ms\ to \rmstwo~\ms. A false--alarm test
yields a 1.1\% probability that the curvature in the residuals of the
single--planet fit is due to random
fluctuations, lending additional credibility to the two--planet
model. However, our data only marginally constrain a two--planet fit
and further monitoring is necessary to fully characterize the
properties of the second companion. To study the effect of stellar
mass on giant planet occurrence, we measure the fraction of stars with
planets in three mass bins comprised of our samples of M Dwarfs,
Solar--mass stars, and intermediate--mass
subgiants. We find a positive correlation between stellar mass and the
occurrence rate of Jovian planets within 2.5~AU. Low--mass K
and M stars have a $\fraclow \pm \fraclowe$\% 
planet occurrence rate compared to $\fracmed \pm \fracmede$\% for
Solar--mass stars and $\frachi \pm \frachie$\% for the higher--mass
subgiants. This result indicates that the former F-- and A--type stars
with $M_* \geq 1.3$~\msun\ in our sample are nearly 5 times more
likely than the M dwarfs to harbor a giant planet. Our analysis shows
that the correlation between Jovian planet occurrence and stellar mass
remains even after accounting for the effects of stellar metallicity. 
\end{abstract}


\section{Introduction}

A planet host star inherits its fundamental characteristics from the
same disk material that forms its 
planets. Studying the relationships between the observed occurrence rate 
of giant planets as a function of the characteristics of
their host stars therefore provides crucial tests of planet formation
theories.  The two most fundamental properties of stars are 
mass and chemical composition. A number of detailed spectroscopic
analyses of nearby stars have revealed a strong correlation between
the metallicity of stars and the likelihood that
they harbor detectable planets \citep{gonzalez97, santos04,
  fischer05b}. This finding can be understood in the context of the
core  accretion model: increasing the metallicity of star/disk system
increases the surface density of solid particulate matter, which leads 
to an enhanced growth rate for protoplanetary cores \citep{ida04b, kornet05,
  ida05a}.

Another way to enhance the surface density of solid material in
the midplane  of the protoplanetary disk is to increase the total mass
of the disk  \citep{ida05b}. If disk masses scale with the
mass of their central stars, then there should be an observed correlation
between planet occurrence and stellar mass. 
The relationship between stellar mass and planet formation rates has
been explored theoretically in the context of the core accretion
model. \citet{laughlin04}  simulated the formation of planets in
disks around low--mass stars and found that the lower surface densities
and longer orbital time scales in the disks around M~dwarfs impede the 
growth of Jupiter--mass planets. By the time the critical core mass
is reached ($\approx10$~\mearth), the supply of disk gas is exhausted
due to accretion onto the central star and
photo--evaporation. \citet{laughlin04} therefore predict an abundance
of ``failed'' gas giants  
with masses comparable to Neptune and a much lower frequency of
Jupiter--mass planets. 

The mass of the central star also 
influences the radial extent of the region in which protoplanetary
cores form. \citet{kennedy07} account for the evolving luminosity of the
central star to model the influence of stellar mass on the
location and size of the core--forming region in circumstellar
disks. Similar to \citet{laughlin04} and \citet{ida05b}, they predict
that M dwarfs should exhibit a deficit of giant
planets. \citet{kennedy07} also predict that the
fraction of stars with Jovian planets should increase with stellar
mass up to a peak of $\approx20$\% for $M_* = 2.5$~\msun.

The prediction of a lower frequency of Jupiter--mass planets around
low--mass M dwarfs is in agreement with the current observational
data. About 300 nearby M dwarfs are currently monitored by various 
Doppler surveys, and only 5
have been discovered to harbor one or more planets
\citep{delfosse98, marcy98, marcy01, endl03, bonfils05b, butler06,
  bonfils07}. Three additional planets have been discovered orbiting
distant ($\sim1$~kpc) low--mass stars by gravitational lensing surveys
\citep{bond04, gould06,   beaulieu06,bennett06}.  

The majority of the Doppler--detected planets around M dwarfs are
significantly smaller than Jupiter, with minimum masses (\msini)
comparable to the masses of Neptune and Uranus
(5--23~\mearth). The sample of low--mass, 
Doppler--detected planets consists of the $22.6 \pm 1.9$~\mearth\
planet around GJ\,436 \citep{butler04,maness07}; the inner, \msini$=
5.9$~\mearth\ planet 
around GJ\,876 \citep{rivera05}; the \msini$= 11$~\mearth\ planet
orbiting GJ\,674 \citep{bonfils07}; and the triple system orbiting
GJ\,581, which consists of planets with minimum masses 16.6~\mearth,
5~\mearth\ and 8~\mearth\ \citep{bonfils05b, udry07}.  
The kinship between these extrasolar ``super Earths'' and the ice
giants in our Solar System was confirmed with the recent discovery
that the Neptune--mass planet around GJ\,436 transits its
central star \citep{gillon07}. The transit light curve provides an
absolute measurement of the planet's mass and radius, suggesting an
internal structure very much like Neptune, with a rocky core
likely surrounded by a thick water layer and thin gaseous envelope. 

To date, there are only two nearby M dwarfs known to harbor
Jupiter--mass companions:  GJ\,876 \citep{marcy98,marcy01,rivera05}
and GJ\,849 \citep{butler06b}. And so far not 
a single ``hot Jupiter'' ($P \lesssim 10$~days) has been discovered
around an M star, in contrast to the $1.2 \pm 0.8$\% occurrence rate
of hot Jupiters around 
Sun--like stars \citep{endl03, butler06}. The relatively
small number of Jovian planets around M dwarfs is not due to a
decreased sensitivity of Doppler techniques for these stars.
For a planet of a given mass and orbital period, the amplitude of a
host star's reflex velocity scales as $K \propto M_*^{-2/3}$, making
planets easier to detect around stars with lower
masses. \citet{butler06b} showed that the frequency of giant 
planets is 2--3 times higher among Solar--mass stars compared to M
dwarfs \citep[See also][]{laws03, bonfils07}. However, the
uncertainty in the estimated planet occurrence rate for M dwarfs is
large due to the small number of target stars and
planet detections. 

Another obstacle that has so far hindered a study of the effects of
stellar mass on planet formation is the limited range of masses
spanned by Doppler--based planet searches. For example, the
Solar--mass, FGK dwarfs 
that  comprise the bulk of the California and 
Carnegie Planet Search (CCPS) span stellar masses from 0.8~\msun\ to
1.2~\msun\ \citep{valenti05, takeda06}. Much more leverage can be
gained by measuring the planet 
occurrence rate around higher mass F-- and A--type stars ($1.3 \leq
M_* \lesssim 3.0$~\msun). Unfortunately, intermediate--mass,
main--sequence stars are poor precision Doppler targets. Stars with  
spectral types earlier than F8 tend to have rotationally broadened
absorption features \citep{donascimento03, galland05}, have fewer
spectral lines due to high surface temperatures, and display a large
amount of excess velocity scatter due to
surface inhomogeneities and pulsation \citep[``jitter'';][]{saar98, wright05}. 

We have addressed the limited range of stellar
masses in Doppler--based planet searches by conducting
a search for planets around stars with masses bracketing those of the
Sun--like stars in the main CCPS sample. At higher stellar masses we are
conducting a search 
for planets around intermediate--mass subgiants at Lick
and Keck Observatories \citep{johnson06b, johnson07}. Subgiants are
evolved stars that have cooler surface temperatures and lower
rotational velocities than their A and F type main--sequence
progenitors ($T_{eff} \approx 5000$~K compared to $T_{eff} > 6000$~K,
and \vsini$ < 5$~\kms\ versus \vsini$ \gg 50$~\kms,
respectively). The spectra of subgiants therefore have an increased
number of narrow absorption lines required for precision Doppler
measurements, making them ideal proxies for A and F type
main--sequence stars. 

At the low--mass end we have been monitoring a sample of 147
late--K through M dwarfs ($M_* \lesssim 0.6$~\msun) as part of the
NASA Keck M Dwarf Planet Survey \citep{butler06, rauscher06}.
The 7--year baseline of our M dwarfs survey, together with the 3--year
duration of our subgiants planet search now provides an excellent
opportunity to measure the relationship between stellar mass and
planet occurrence for wide range of stellar masses and comparable
detection characteristics. Here, we
report the detection of a Jupiter--mass planet in a 
\pyearsA~yr orbit around the M3.5 dwarf \starA. We present the
stellar characteristics of the host star in \S~\ref{stellar}. In
\S~\ref{orbit} we present our observations and orbit solution,
including an assessment of a possible second Jovian companion in the
system. In \S~\ref{massplanet} we incorporate this latest M dwarf
planet detection into a detailed analysis of the 
relationship between stellar mass and the occurrence rate of giant
planets. We conclude in \S~\ref{discussion} with a summary and a brief
discussion of our results.

\section{Properties of \starA}
\label{stellar}

\starA\ (LHS\,2037, L\,675-081), is a nearby M3.5 dwarf among
the 147 low--mass stars in the NASA Keck M~Dwarf Survey. The Simbad
database lists an apparent magnitude $V = 13.0$. However, this
magnitude is 
$\sim1$~mag fainter than the values listed in several other catalogs:
The Gliese--Jahreiss Catalog \citep[GJ Catalog;][]{gj91} gives $V = 
\vmagA$, the Carlsberg Meridian Catalogs list $V = 12.03$, and 
the Catalogue of Stellar Spectral Classifications lists $V = 11.98$
\citep{skiff05}. We adopt the mean of these measurements, $V = \vmagA$.
The GJ Catalog also lists $B-V = \bvA$ and a trigonometric parallax
$\pi = 101.3 \pm 27$~mas. The GJ 
Catalog color agrees well with the color measured by
\citet[$B-V = 1.53$;][]{reid02}, and the listed parallax is consistent
with the value 
measured by \citet{woolley70} ($\pi = 116 \pm 12$~mas). We adopt the mean
of the two trigonometric parallax measurements, $\pi = \parallaxA \pm
\parallaxeA$~mas. This value is 1-$\sigma$ larger (based on the
quadrature--sum of the uncertainties) than
the ``resulting parallax'' $\pi = 87 \pm 17$~mas listed in the the GJ 
Catalog, which 
is estimated from the star's broadband colors and spectral type. Our
adopted trigonometric parallax yields a distance of $\dA \pm 1.7$~pc
and an absolute visual magnitude $M_V = \mvA$.  

\begin{figure}[t!]
\epsscale{1}
\plotone{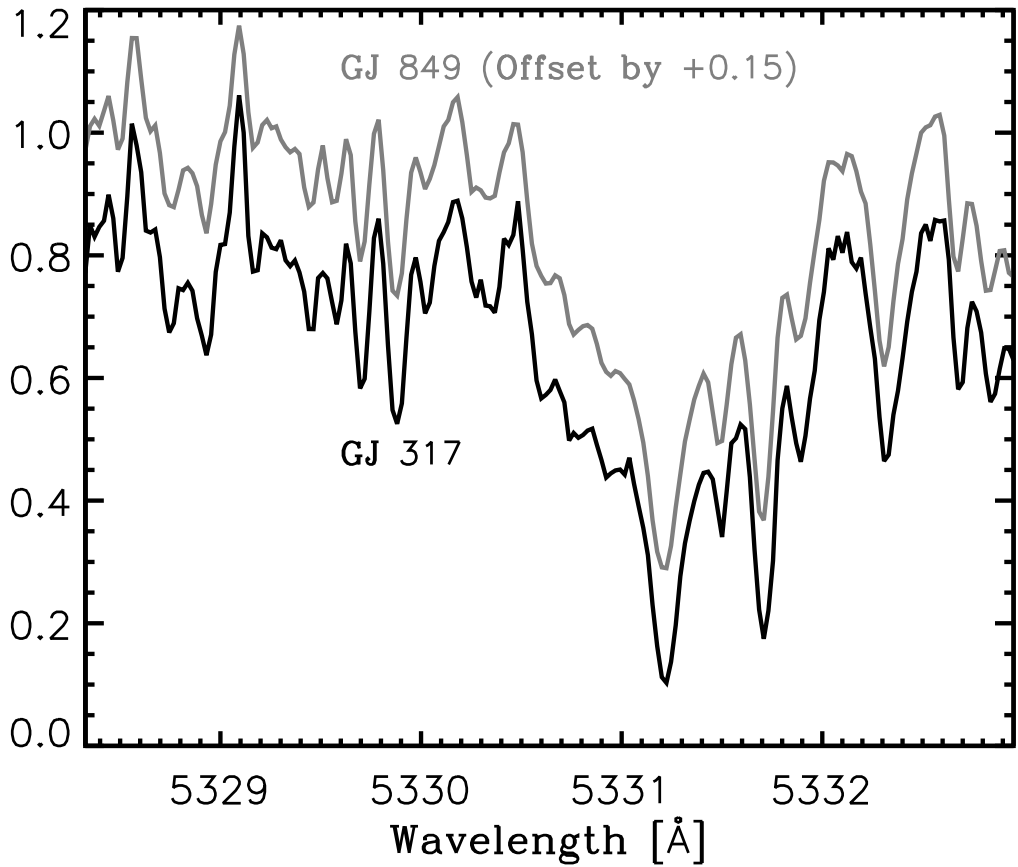}
\figcaption{\footnotesize Comparison of the spectra of \starA\ (black line) and
  GJ\,849 (gray line), which   has a similar $B-V$ color, spectral
  type and metallicity (\fe$=+0.16 \pm 0.2$). The spectrum of
  GJ\,849 has been shifted up by 0.15 for clarity. Since the spectral
  lines of the two stars have very similar depths, we conclude that
  \starA\ cannot have a metallicity much different than that of
  GJ\,849. \starA\ is therefore not a metal--poor subdwarf, despite
  its apparent location 2~mag below the mean Hipparcos main sequence.
  \label{spec_compare}}
\end{figure}

For its $B-V$ color \starA\ is unusually faint, lying
$\approx1.8$~mag below the mean Hipparcos main sequence as defined by 
\citet{wright04}. Its location in the H--R diagram suggests that
\starA\ is an extremely metal--poor subdwarf. However, the star's
2MASS infrared magnitudes together with the K--band
photometric metallicity--luminosity calibration of \citet{bonfils05a}
 suggest that \starA\ has a metallicity
consistent with solar, \fe$ = \feA \pm 0.2$. This metallicity is much
higher than the value expected given the star's position in the H--R
diagram. 

We compared our
iodine--free template spectrum of \starA\ to a template spectrum of a
similar M dwarf, GJ\,849 (M3.5, $V = 10.4$, $B-V = 1.52$, [Fe/H]$ =
+0.16 \pm 0.2$, \citet{butler06b}). Figure~\ref{spec_compare} shows a
comparison of a small portion of the template spectra, which shows
that these two stars are remarkably similar. Since the line depths 
are nearly identical we can safely conclude that \starA\ is not
a metal--deficient subdwarf. As an additional check, we
compared the 2MASS $J-K$ and $H-K$
colors of GJ\,317 to those of GJ\,849. The infrared colors of the two
stars agree to within 0.001 and 0.006 mag, respectively, again
suggesting very similar spectral characteristics.

We can imagine several possible explanations for the abnormal location of
\starA\ in the H--R diagram. The first possibility is that the
parallaxes listed in the literature are systematically
incorrect, and \starA\ is actually significantly further away than
the reported parallaxes suggest. Similarly, the star's reported
apparent $V$ magniude may be too faint. Another possibility is that
the star's $B-V$ color is too blue by $\sim0.1$~mag. Finally, the star
may be obscured by 1.5~mag of gray extinction. We are planning
follow--up photometric monitoring during the next observing season to
further investigate the latter three scenarios.  

We used the 2MASS infrared photometry of \starA, together with the
K--band mass--luminosity calibration of \citet{delfosse00} to estimate
a stellar mass $M_* = \kmass \pm 0.04$~\msun. This mass estimate agrees
well with the $\vmass \pm 0.05$~\msun\ value predicted by the V--band
mass--luminosity relationship of \citet{bonfils05a}. We adopt the mean
of these two estimates,  $M_* = \mstarA$~\msun. The properties of
\starA\ are summarized in Table~\ref{star_props}.   

\begin{figure}[t!]
\epsscale{0.6}
\plotone{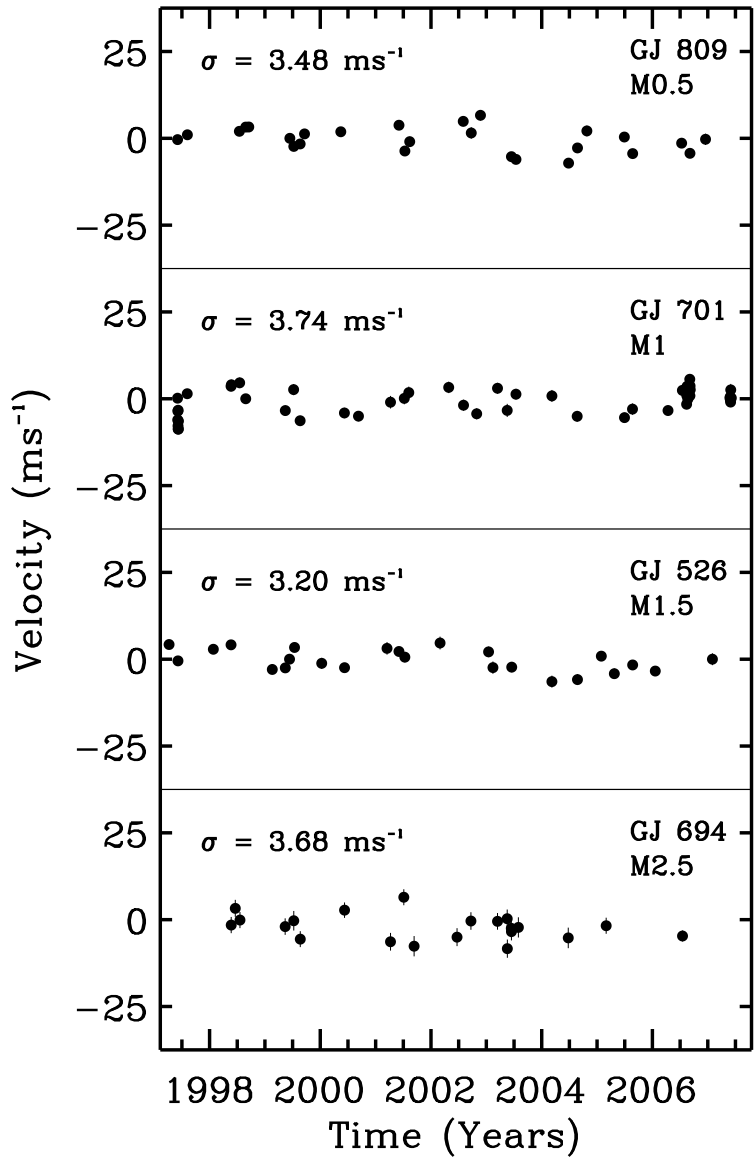}
\figcaption{\footnotesize Radial velocity time series for four stable
  M~dwarfs in our Keck Doppler survey demonstrating a Doppler precision
  of 3--4~\ms\ over the past 8--9 years. \label{keck_stable}} 
\end{figure}

\section{Observations and Orbital Solutions}
\label{orbit}

We have been monitoring \starA\ with the Keck I 10\,m telescope
for 7.4 years.  We obtained high--resolution spectra using the HIRES
echelle spectrometer \citep{vogt94} with an iodine cell mounted
directly in front of the entrance slit. The dense
set of molecular absorption lines provide a robust wavelength
fiducial, as well as information about the shape of the spectrometer
instrumental profile 
\citep{marcy92b, valenti95}. The Doppler shift is measured from each 
star--plus--iodine observation using the modeling procedure described
by \citet{butler96}. Figure~\ref{keck_stable} shows velocity
measurements of four stable M~Dwarfs, demonstrating our long--term
Doppler precision of 3--4~\ms.

\begin{figure}[t!]
\epsscale{1}
\plotone{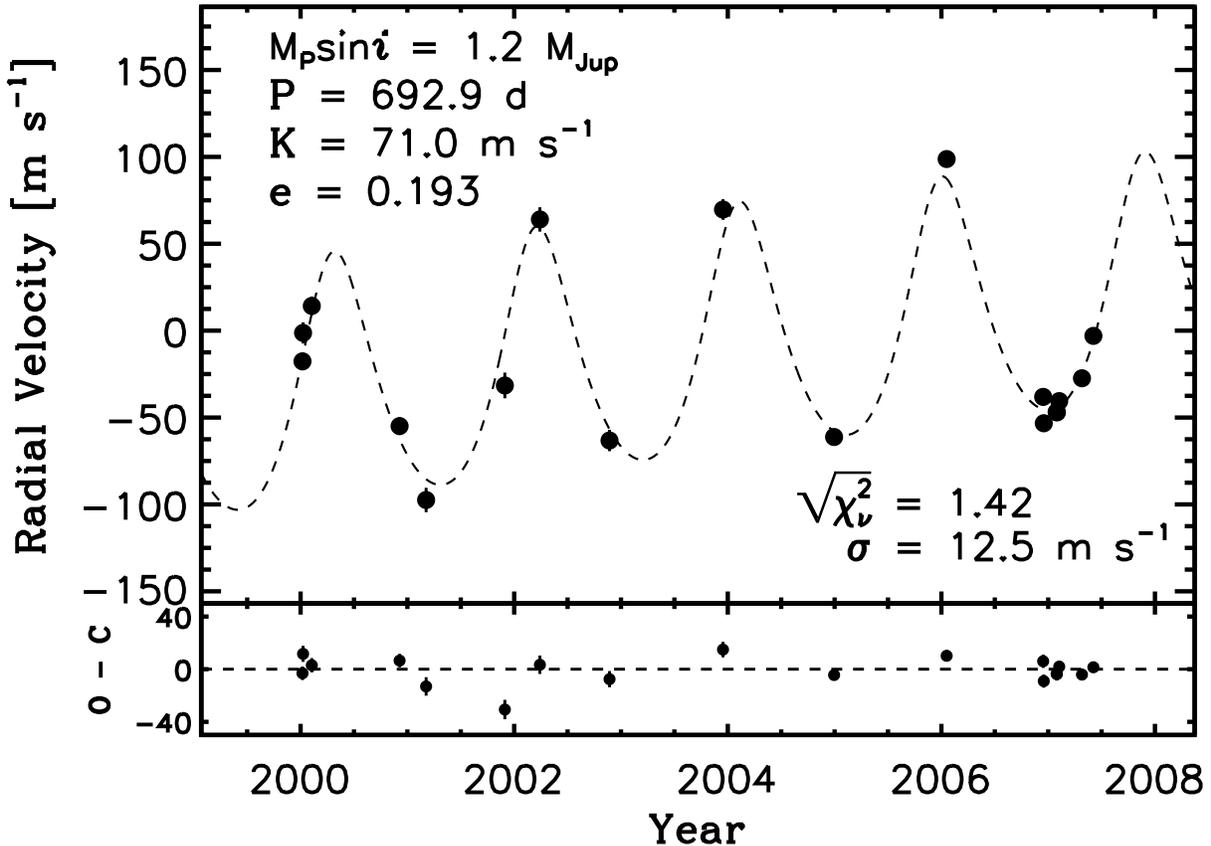}
\figcaption{\footnotesize Our radial velocity time series for \starA, based on
  observations from Keck observatory. The dashed line shows the
  best--fit Keplerian with an additional linear trend of $\trendA \pm
  \trendeA$~\ms\,year$^{-1}$. The rms scatter of the residuals (bottom panel) is much
  larger than the predicted 6~\ms\ of scatter due to mearurement
  uncertainties and stellar jitter. \label{orbit1}}
\end{figure}

A total of \nobsA\ Doppler measurements of \starA\ are listed in
Table~\ref{velgl317} and shown in 
Figure~\ref{orbit1}. We carried out a search for the best--fit
Keplerian orbital solution using a nonlinear, least--squares
algorithm. In the fitting procedure, each velocity measurement is
assigned a weight constructed 
from the quadrature sum of the internal measurement uncertainty and a
stellar ``jitter'' term. The internal uncertainty is
the weighted 
standard deviation of the mean velocity measured from each of the
$\sim700$ 2-\AA\ chunks in each echelle spectrum. 
The stellar jitter term accounts for noise in excess of the internal
uncertainties due to astrophysical sources such as rotational
modulation of stellar surface features and stellar pulsation
\citep{saar98, wright05}. For Sun--like 
dwarfs, the stellar jitter can 
be accurately predicted based on the chromospheric diagnostic \rphk,
given by the ratio of the CaII\,H\&K flux to the stellar UV
continuum, along with luminosity and $B-V$ color. However, \rphk\
is not calibrated for faint M dwarfs because stars with such cool
atmospheres lack a well defined continuum due to molecular line
blanketing. We therefore estimate the stellar jitter of the M dwarfs
in our sample by assuming a constant value for all stars 
with $B-V > 1.3$. We selected the 133 M stars in our sample with 12 or
more observations and measured the jitter using the
prescription of \citet{wright05}. For HIRES observations made prior to
the 2004 CCD upgrade we derive a median jitter of 3.5~\ms, and 
2.0~\ms\ for post--upgrade observations.

Our search for a best fitting Keplerian to the \nobsA\
observations of \starA\ reveals a minimum in \chisq\ for an orbital
period near $P = 690$~days. However, the root--mean--squared (rms) residual
to this best fitting, single--planet model is \rmso~\ms, which is much
higher than the expected scatter of $\approx 6$~\ms\ from the internal
errors and jitter. The resulting reduced \chisq$=\chio$ indicates the
single--planet model is inadequate.

We attempted to improve on the single--Keplerian model by adding a
variable linear trend. The existence of a trend
in the velocities implies an 
additional, long--period orbital companion in the system. The dashed
line in Figure~\ref{orbit1} shows the best fitting Keplerian with a
linear trend of \trendA$\pm$\trendeA~\ms\,yr$^{-1}$. The inclusion of
a trend reduces the rms from \rmso~\ms\ to \rmstr~\ms, and
decreases \chisq\ from \chio\ to \chitr, after accounting for the extra
free parameter. The Keplerian parameters of this fit are listed in
Table~\ref{kep_pars}, along with their estimated uncertainties. The
parameter uncertainties were derived using a Monte Carlo method
\citep[e.g.][]{marcy05b}. Our best--fit orbital solution yields a
\pyearsA\ year period,  velocity semi--amplitude $K = \kA$~\ms, and
eccentricity  $\eA \pm \eeA$. Using our adopted stellar mass $M_* =
\mstarA$~\msun, we calculate a minimum planet mass $M_P\sin{i} =
\msiniA$~\mjup\ and semimajor axis $a = \arelA$~AU. 

\subsection{The Two--Planet Model}

\begin{figure}[t!]
\epsscale{1}
\plotone{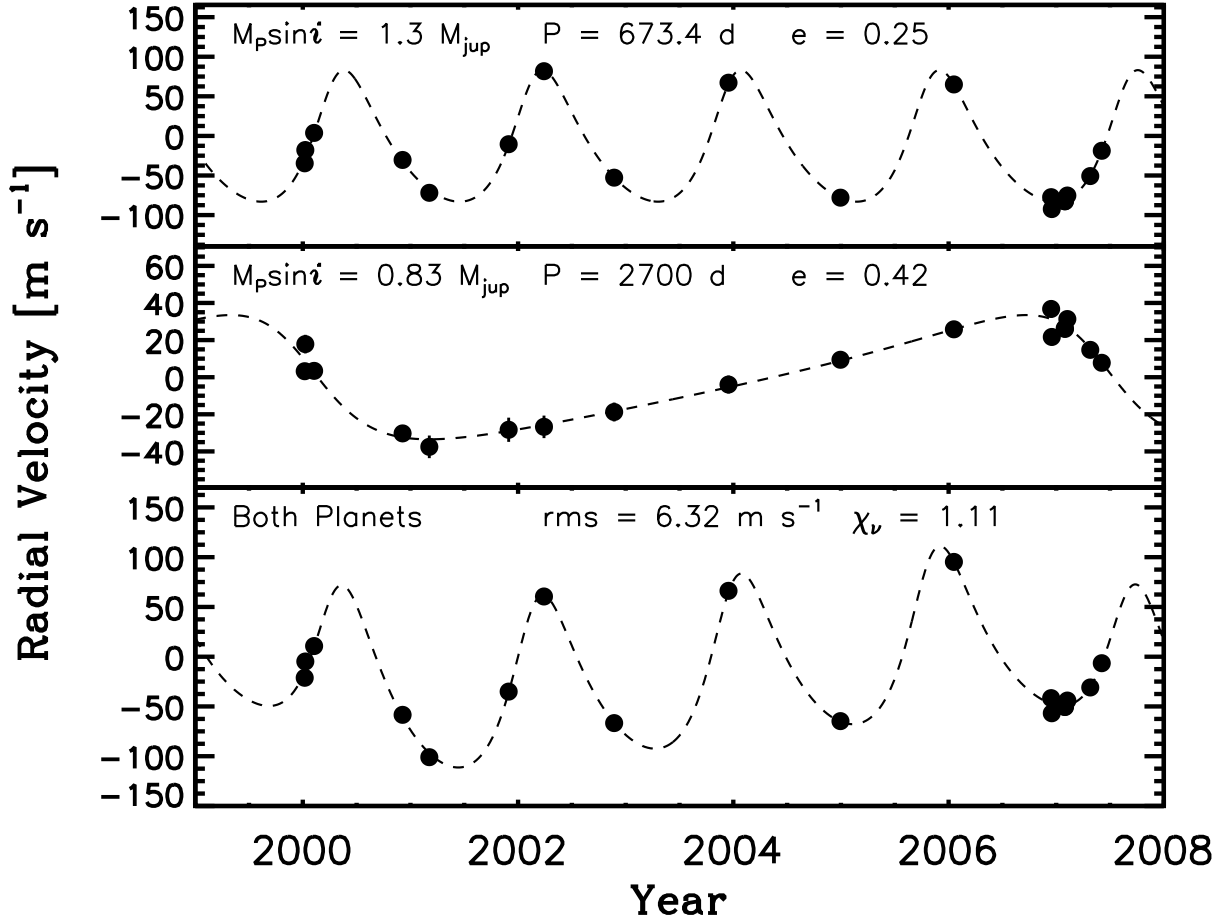}
\figcaption{\footnotesize The best fitting two--planet model. The top
  two pannels show the single--planet fits with the other component
  removed. The addition of an outer  
  planet with a period $P \approx 2700$~days decreases the rms scatter
  of the residuals from \rmstr~\ms\ to \rmstwo~\ms, and the reduced
  \chisq\ from \chitr\ to \chitwo. However, the duration and
  time--sampling our observations only 
  marginally constrain the 11 free parameters of the two--planet fit. 
  \label{orbit2}} 
\end{figure}

In light of the success of the Keplerian--plus--trend model, we
performed a search for a double--Keplerian fit. The 
best fitting, two--planet model is shown in Figure \ref{orbit2}, with
orbital periods of \pBB\ and \pBC\ days, respectively. The minimum
mass of the inner planet changes from \msiniA~\mjup\ to
\msiniBB~\mjup, compared the Keplerian--plus--trend model. The rms
scatter drops from \rmstr~\ms\ to \rmstwo~\ms\, with a resulting
decrease in \chisq\ from \chitr\ to \chitwo.

\begin{figure}[t!]
\epsscale{1}
\plotone{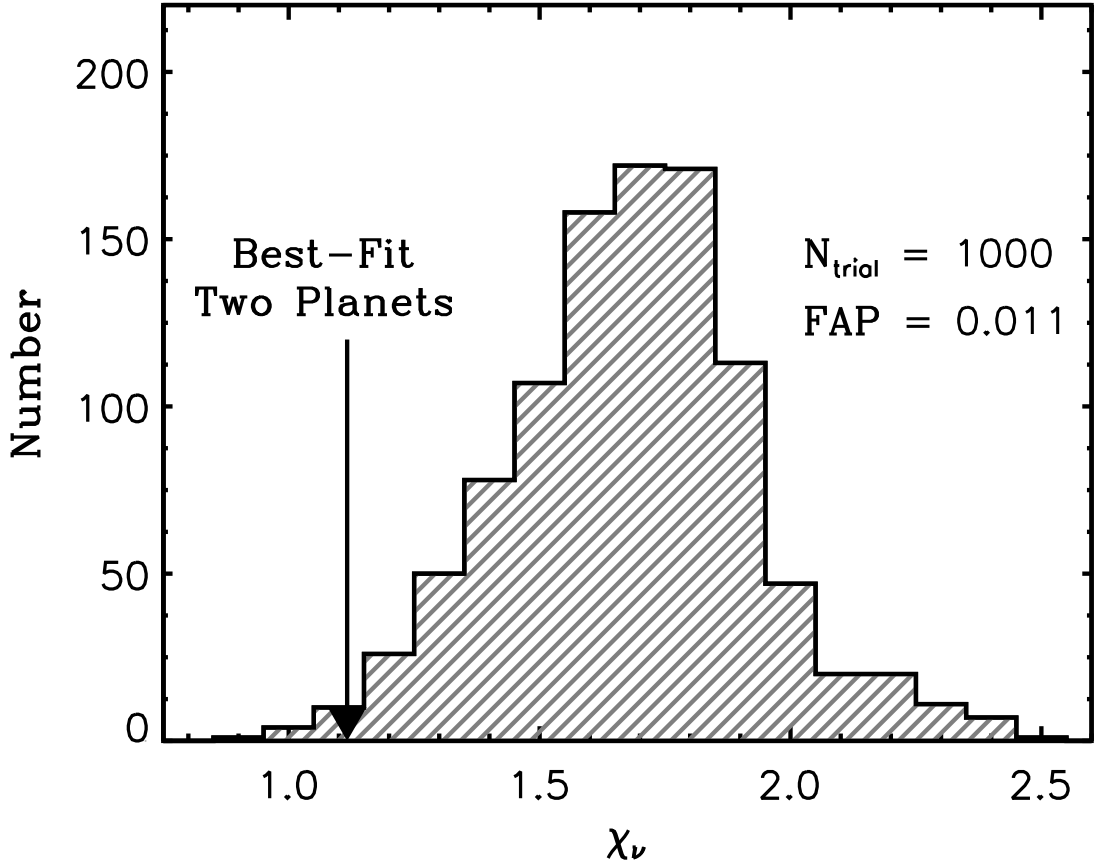}
\figcaption{\footnotesize Empirical assessment of the false--alarm
  probability of the two--planet model. The observed velocities yield a
  double--planet orbital solution with \chisq$ = \chitwo$ (arrow).
  The histogram shows the distribution of \chisq\ obtained from
  single--Keplerian fits to the mock velocity sets, with a variable
  linear trend. The mock data sets 
  are generated by adding the scrambled residuals from a
  single--Keplerian fit to the observed velocities. 11 of the 1000
  mock data sets produced orbital fits with \chisq\ lower than the 
  original time series, indicating a FAP of 0.011 that random
  fluctuations are a viable explanation for the quality of the
  two--Keplerian fit to the observations. Additional observations are
  required to fully characterize the outer companion and reduce the
  FAP to below 0.01. \label{fap}} 
\end{figure}

While the quality of the double--planet fit is encouraging,
it is still possible that random variability and jitter could conspire
to produce a false curvature in our sparse series of
measurements. If this is the case, then the second Keplerian and its
associated 4 additional parameters are an unjustified embellishment on
the single--Keplerian--plus--trend model. We tested this null
hypothesis by employing the 
scrambled--residuals false--alarm test described by
\citet{wright07}. We first subtract the single--Keplerian--plus--trend
model  from the data and adopt the residuals as random variates to
construct a large number of mock velocity time series. Each mock time
series is constructed by scrambling the residuals, with replacement,
using a pseudo--random number generator. The scrambled residuals are
then added back to original data to generate a mock data set. For each 
mock time series we perform a full search for a best--fit
Keplerian--plus--trend orbital solution. If the curvature in the
original data is an artifact of random noise, then many of the mock
data sets should produce lower \chisq\ than the two--planet fit to the
original data. However, if a second, resolved Keplerian signal is
present in the observations, then the mock data sets should produce
fits that are much worse on average than the original velocities. 

The results of our test are shown in Figure~\ref{fap}. We generated
1000 mock data sets and found that only 11 produced \chisq\ equal to
or lower than the original velocities, resulting in an false--alarm
probability FAP$=0.011$. The 1.1\% FAP is promising, but we
are nonetheless cautious about the validity of this 
two--planet model. Given the relatively low amplitude of the second signal and
long period of the two candidate planets, our \nobsA\ measurements
provide only 12 independent data points. The two--planet model
is therefore only marginally constrained by our present data. Further, the
duration of our observations only covers one orbit cycle of the
putative outer companion. At this point we can only
confirm the inner planet, and additional observations are required to
verify the exact orbital properties of \starA\,c.

\section{The Stellar Mass--Exoplanet Correlation}
\label{massplanet}

\subsection{Sample Selection}

The mass range encompassed by the California and Carnegie
Planet Search target stars provides a large, uniform sample of targets
that are ideal for evaluating 
the effect of stellar mass on the occurrence of Jovian planets. This
analysis can be accomplished by simply measuring the fraction of stars 
with planets in three mass bins: the low--mass M and late K
dwarfs with $M_* < 0.7$~\msun, the Sun--like FGK stars with  $0.7 \leq
M_* < 1.3$~\msun, and the intermediate--mass subgiants with $1.3 \leq M_* \leq
1.9$~\msun. 

We were careful to select planets that could be reasonably
detected around stars with a wide range of masses, and for surveys
with different time baselines.  
Our first requirement is that stars have at least 8 observations 
spanning enough time to search for planets out to a
a distance of 2.5~AU. Using Kepler's third law,
the orbital period scales as $P \propto M_*^{-1/2}$ for a fixed
semimajor axis.  An orbital distance of 2.5~AU roughly corresponds to
periods equal to the 3 year duration of the 
subgiants planet search, as well as the $\sim7$ year duration of the
NASA Keck M Dwarfs Survey. Next, we exclude 
stars selected as part of the metallicity--biased Next 2000 Stars
(N2K) survey \citep{fischer05a}. Metallicity is an established tracer 
of giant planets, and we want to isolate the effects of stellar
abundance from the effects of stellar mass. We also excluded stars
that were added to the CCPS target list after a planet had already
been announced by another group. 

We included only planet
detections that induce a velocity semi--amplitude, $K$, large enough
to be detected around stars in all three mass bins. For a
planet of a given minimum mass and orbital period, the host star's
velocity amplitude scales as $K \propto M_*^{-2/3}$. The larger masses of
the subgiants therefore limit the minimum planet mass that can be
included in our analysis. \citet{johnson07} find that their Doppler
measurements of subgiants yield a typical precision of $\sim
6$~\ms. We therefore include in our analysis only planets with minimum
masses \msini$ \geq 0.8$~\mjup. A planet with this minimum mass in a
$P=3$~year orbit would represent a 2-$\sigma$ detection around a
1.6~\msun subgiant, and would be much easier to detect around a
lower--mass star.

Our choice of such a large minimum planet mass is conservative
since lower--mass planets would be readily detectable in shorter
orbits. However, restricting our analysis to planets with masses
greater than 0.8~\mjup\ also accomplishes a separate goal. We wish to 
test the effects of stellar mass on the formation of \emph{giant}
planets. The models of \citet{ida05a} and \citet{laughlin04}
demonstrate that the formation of planets like Jupiter is a threshold
process limited by the growth rate of the rocky, embryonic cores. The
core--growth rate is in turn limited by the supply of solid material
at the midplane of the protoplanetary disk, which should scale with the 
mass of the central star.

\subsection{Jovian Planet Occurrence}

\begin{figure}[t!]
\epsscale{1}
\plotone{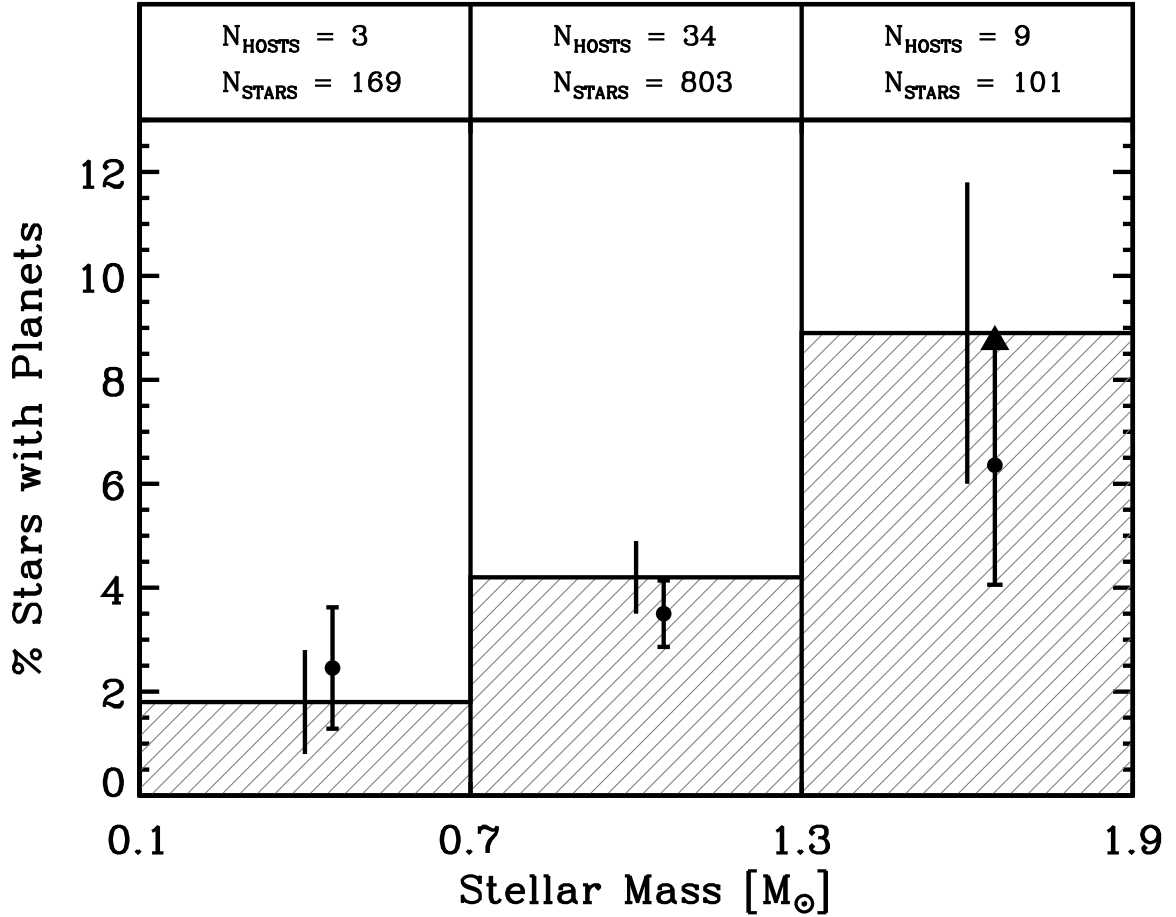}
\figcaption{Histogram illustrating the rising percentage of stars with
  detectable planets as a function of stellar mass. The stars selected
  for each mass bin have 8 or more observations providing
  detectibility of planets with masses \msini$ \geq 0.8$~\mjup\
  out to $a = 2.5$~AU.
  After correcting the measured percentages in each
  mass bin for the effects of stellar metallicity, the rising trend
  is slightly diminished (filled circles, offset to the right by
  +0.05~\msun\ for clarity). However, the high--mass bin is
  uncorrected for the decreased sensitivity of Doppler measurements
  of higher--mass subgiants compared to lower--mass stars (cf
  \S~\ref{metcorr}). Our measured occurrence rate for high--mass
  stars therefore represents 
  a lower limit unlike the Solar--mass and low--mass bins.
  The error bars on each bin are from
  Poisson statistics and the numbers above each bin compare the number
  of stars with planets $N_{HOSTS}$ to the total number of stars in each
  bin $N_{STARS}$. 
  \label{mass_hist}}  
\end{figure}

The results of our analysis are displayed in
Figure~\ref{mass_hist} and are summarized here.
For the low--mass bin, we selected stars with masses $M_* < 0.7$~\msun.
Of the 147 low--mass stars in the NASA M Dwarf Survey, 130 meet our 
selection criteria. These stars
have masses estimated using the \citet{delfosse00} K--band
mass--luminosity calibration. We also calculated the masses of all stars
with $B-V > 1.3$ using the Padova stellar interior models
\citep{girardi02}. This analysis revealed an additional 39 stars in
the mass range $0.6 < M_* \leq 0.7$~\msun. Of these 169 total stars,
only 3 are known to harbor at least one giant planet: GJ\,876, GJ\,849
and GJ\,317. The resulting planet occurrence rate for these low--mass
stars is $\fraclow \pm \fraclowe$\%.

For the Solar--mass bin we selected stars from the CCPS Lick, Keck and
Anglo--Australian Observatory surveys with masses in the range
$0.7 < M_* \leq 1.3$~\msun. A total of 803 stars met our stellar mass and
planet--detectability criteria. Of these stars, 34 harbor at least one
giant planet, resulting in a planet occurrence rate of $\fracmed \pm
\fracmede$\%. 

For the high--mass bin, we used the intermediate--mass subgiants in
the \citet{johnson06b} sample, of which 68  meet our selection
criteria and have masses in the range $1.3 < M_* \leq 1.9$~\msun\
according to the \citet{girardi02} stellar interior models. We found an
additional 33 intermediate--mass subgiants in the regular CCPS Keck
and Lick samples, for a total of 101 stars. There are 6 
published planetary companions among these stars. An additional 3
companions will be 
announced in a future publication pending follow--up observations at
Lick Observatory in the upcoming observing seasons (Johnson et
al. 2007b, in preparation). While these additional planet candidates
lack sufficient 
phase coverage to publish at this time, their signals currently have
false--alarm probabilities FAP$ < 0.01$\footnote{We have analyzed the
  velocities for our entire sample of M dwarfs and find no candidate
  signals due to Jovian--mass planets with FAP$< 0.01$. Similarly,
  excluding the metal--rich N2K targets, none of the FGK stars in the
  CCPS samples has an unpublished candidate with \msini$ \geq
  0.8$~\mjup\ within 2.5~AU.}. Including these strong candidates
results in 9 giant planets among 101  intermediate--mass subgiants,
for an occurrence rate of $\frachi \pm \frachie$\%. 

\subsection{The Effects of Metallicity}
\label{metcorr}

The occurrence of detectable planets has previously been shown to
correlate with stellar metallicity
\citep{gonzalez97, santos04, fischer05b}. It is therefore important to
test whether 
the correlation between stellar mass and planet occurrence
shown in Figure~\ref{mass_hist} is real, or merely an effect of an
underlying metallicity bias. 

We searched for such a bias in our sample by measuring the median
metallicity in each mass bin. Since the Solar--mass stars and
intermediate--mass subgiants were analyzed using the same LTE
spectral synthesis code (SME), our results from these two mass regimes
can be directly compared. For the Solar--mass bin we find a
median metallicity of [Fe/H]$=+0.042$. The median metallicity of the
subgiants is slightly higher at [Fe/H]$=+0.075$.
In order to compare the expected occurrence rate in each of these mass
bins based on metallicity alone, we employ the relationship between
metallicity and planet probability from \citet{fischer05b},

\begin{equation}
\label{planetprob}
P(\rm planet) = 0.03 \times 10^{2[Fe/H]}.
\end{equation}

Given the median metallicity of the Solar--mass stars,
Equation~\ref{planetprob} predicts a 3.6\% probability of finding a
detectable planet. This is consistent within errors 
with the $\fracmed \pm \fracmede$\% probability that we measure. 
For the subgiants, the expected probability is 4.2\%, which is a
factor of 2 lower than the $\frachi \pm \frachie$\% we
measure. This enhanced planet occurrence around our high--mass stars
is significant at the 1.6-$\sigma$ level. 

The correlation between stellar mass and exoplanets in
Figure~\ref{mass_hist} is most apparent when comparing the bins at
either mass extreme. The 
high--mass stars in our sample appear to be planet--enriched
by a factor 5 compared to the low--mass M dwarfs, which
represents a 2.3-$\sigma$ result. However, it is currently very
difficult to derive spectroscopic abundances for M
dwarfs\footnote{See, however, \cite{bean06} for spectroscopically
  derived metallicities of 3 M dwarf planet host stars.}. Could
it be that there is a systematic metallicity bias among our low--mass
stars? 
\clearpage

A common argument for such a metallicity bias among M dwarfs
is related to their age \citep[e.g.][]{bonfils05b}. According to this
argument, M dwarfs have lifetimes longer than the age of the Galaxy
and are therefore 
more likely than higher--mass stars to have formed in the distant past
when the Galaxy was metal poor. This argument is based on the
assumption that there is a well--defined age--metallicity relationship
in the Galaxy. However, studies of Galactic chemical evolution 
have shown that no well--defined correlation between age and
metallicity exists in the thin Galactic disk. Analyses of the local
stellar population show that stars 
have a wide range of abundances at ages from 1 to 10
Gyr, with a 0.2 dex scatter in [Fe/H] over this age range
\citep{edvardsson93, 
  nordstrom04, takeda06}. Similarly, open clusters with comparable
ages also exhibit a wide range of metallicities, with 3 to 6 Gyr
clusters differing by as much as 0.82 dex \citep[e.g.][]{chen03}. 
Further, late G-- and all K--type stars also have
lifetimes comparable to the age of the Galaxy. Therefore, any
metallicity bias present in our sample of M dwarfs should also be
present in a volume--limited sample of G and K dwarfs. We therefore
find no compelling reason to suspect that the abundances of M dwarfs
should deviate significantly from the G and K stars in the Solar
neighborhood. 

Our sample of M dwarfs within 15~pc of the Sun should trace the metal
abundances of the other stars in the immediate Solar
neighborhood. The volume--limited sample of \citet{fischer05b} has a 
mean metallicity of [Fe/H]$ = -0.05$. This is roughly the abundance we
expect for our sample M dwarfs. We stress that, contrary
to the argument above,  our sample of M dwarfs is not metal--poor
compared to other nearby stars. Rather it is the 
Solar--mass stars and subgiants among the planet search targets that
are slightly metal rich compared to a volume--limited sample. This
effect has been previously noted by 
\citet{fischer05b} and is related to the fact that in a given range of
$B-V$ colors metal--rich stars are brighter and more likely to be
included in the samples of most Doppler--based
planet searches \citep[see also][for a more detailed discussion.]{marcy05b}.

Based on our estimate for the mean metallicity of our sample of M
dwarfs ([Fe/H]$ = -0.05$), Equation~\ref{planetprob} predicts a 2.4\%
planet occurrence rate for the stars in our low--mass bin. This
prediction is slightly higher than our measured occurrence rate of
$\fraclow \pm \fraclowe$\%, but the two values agree within errors. 

We can use the probabilities predicted by Equation~\ref{planetprob} to
correct our measured occurrence rates for the effects of stellar
metallicity. Each corrected planet fraction, $f^\prime_i$, is
constructed as

\begin{equation}
f^{\prime}_i = (p_\sun/p_i) \times f_i,
\end{equation}

\noindent where $f_i$ is the original planet
fraction, $p_\sun = 0.03$ is the expected probability for solar
metallicity, and $p_i$ is the predicted probability from 
Equation~\ref{planetprob}. Similarly, the corrected uncertainty in
each bin is 

\begin{equation}
\sigma^{\prime}_i = \sqrt{p_\sun/p_i} \times \sigma_i .
\end{equation}

\noindent The
filled circles in Figure~\ref{mass_hist} show the  corrected values
and uncertainties of the planet occurrence rate in each mass bin. After
correcting for metallicity, the contrast between 
the low--mass and high--mass bins is diminished. However, evidence for
an increasing trend remains: the A--type stars have a factor of 2.5
higher planet occurrence rate than the M dwarfs.  

It is important to note that the corrected occurrence rates in
Figure~\ref{mass_hist} do not account for the decreased 
sensitivity of our Doppler survey of subgiants compared to the
surveys of lower--mass stars. Our requirement of 8 or more
observations was exceeded in most cases by the stars in the
Solar--mass and low--mass bins. The Solar--mass stars had on average
21 observations each, and the M dwarfs averaged 16 observations. 
On the other hand, the subgiants had an 
average of only 10 observations each. Also, whereas our typical
Doppler precision is $\lesssim$2~\ms\  for Solar--mass
stars \citep{marcy05b, johnson06a} and 3--4~\ms\ for M dwarfs
(Figure~\ref{keck_stable}), the subgiants typically have a precision
of 4--7~\ms\ due to their larger jitter \citep{johnson07}. The
sensitivity of our 
measurements of subgiants is further diminished since the amplitude of
a star's reflex motion caused by a planet with a given mass and orbital period
scales as $K \propto M_*^{-2/3}$. We therefore 
conclude that the measured planet occurrence rate for the stars in our
high--mass bin represents a lower limit rather than an absolute
measurement of the true fraction of subgiants with planets (as
indicated by the upward--facing arrow in Figure~\ref{mass_hist}), even
after correcting for metallicity. 

\section{Summary and Discussion}
\label{discussion}

We present the detection of a Jupiter--mass planet
orbiting the M3.5 dwarf \starA. This detection marks the sixth
M~dwarf known to harbor at least one Doppler--detected  planet,
together with GJ\,876 
\citep{marcy98, delfosse98, marcy01, rivera05}, GJ\,436
\citep{butler04}, GJ\,581 \citep{bonfils05b, udry07}, GJ\,849
\citep{butler06b} and GJ\,674 \citep{bonfils07}. \starA\ is the only
third M dwarf out of roughly 300 surveyed known to harbor a Jovian
planet. In addition to our firm detection of a Jovian planet with $P =
\pA$~days, we also detect evidence of a possible second Jovian planet
in the system near $P \approx 2700$~days. However, additional
monitoring is required to fully characterize the orbit of the second
planet.   

Multi--planet systems appear to be relatively common among M dwarfs
compared to Sun--like stars. All M stars with one Jovian planet show
evidence of a second companion. GJ\,876 has a pair of Jupiter--mass
planets in a 2:1 mean motion resonance, along with an inner ``super
Earth'' \citep{marcy01, rivera05}. GJ\,849 has a long--period Jovian
planet with a linear trend 
\citep{butler06b}. Of the 3 M dwarfs with Neptune--mass planets, two
have multiple planets or evidence of an additional companion: GJ\,581
harbors 3 low--mass planets \citep{udry07}, and GJ\,436 has a linear
trend \citep{maness07}. Only GJ\,674 appears to be in a
single--planet system. From the first 6 planet
detections around low--mass stars, it appears as though M dwarfs have
an 80\% occurrence rate of multi--planet systems, compared to the 30\%
rate measured for FGK stars \citep{wright07}. 

The high frequency of multi--component detections around M dwarfs may
be in part due to the increased detectability of planets around
low--mass stars. This is likely the case for the triple Neptune system
around GJ\,581. The ``c'' component in that system would have induced
a velocity amplitude of only 1.4~\ms\ if it orbited a 1~\msun\
star. However, the enhanced detectability of planets around M stars
shouldn't matter as much for the systems containing Jovian planets
since all of the giant planets orbiting M stars would be detectable
around Solar--mass stars.
 
We studied the relationship between stellar mass and the occurrence rate
of giant planets by combining our sample of M dwarfs with our
samples of Solar--mass FGK stars and intermediate--mass subgiants. By
measuring the fraction of stars with planets in 
three stellar mass bins we find that the frequency of planets with
$M_* > 0.8$~\mjup\ within $a < 2.5$~AU increases with
stellar mass (Figure~\ref{mass_hist}). The retired A--type stars in
our sample are nearly 5 times more likely than M dwarfs to harbor a
giant planet. This important result establishes stellar mass as an
additional sign post for exoplanets, along with metallicity. Just as
metallicity informs the target selection 
of searches for short--period planets \citep[e.g.][]{fischer05a},
stellar mass will be an important factor in the target selection of
future high--contrast direct imaging surveys. While the lower
luminosities of M dwarfs provide favorable contrast ratios that
facilitate the detection of thermal emission from young giant planets,
our results show that A--type stars are far more likely to harbor such
planets. 

In order to understand the role of stellar mass
on planet formation, it is important to disentangle the known effects of 
stellar metallicity from our stellar sample. After correcting for
metallicity in each mass bin, the slope of the trend in
Figure~\ref{mass_hist} is 
slightly diminished. However, the factor of 2.5 increase in planet
occurrence around A stars compared to M dwarfs remains significant,
especially 
considering that the high--mass bin is uncorrected for the relatively
lower detection sensitivity of the subgiants planet search. Our
results therefore confirm the prediction of the core accretion
model that higher mass stars should form Jovian planets more
efficiently \citep{laughlin04, ida05b, kennedy07}. 

Our discovery confirms and expands upon the recent results of
\citet{lovis07}, who are searching planets around intermediate--mass 
K giants. By focusing on stars in open clusters, they are able to
control for the age and metallicity of their sample and determine
accurate stellar masses. Because of the larger stellar radii and
lower precision (jitter~$> 15$~\ms) of
K giants, they focused on planets with \msini$ \geq 5$~\mjup\
and $0.5 \geq a \geq 2.5$~AU. \citet{lovis07} 
find that the occurrence rate of ``super--Jupiters'' and
brown dwarfs within this mass and semimajor axis range increases with
stellar mass, rising from 0\% for M dwarfs to 2.5\% for evolved A
stars. 

The high--mass sample of stars analyzed by \citet{lovis07} contained
primarily stars with masses $M_* \gtrsim 2.0$~\msun. In order to avoid
confusion with horizontal--branch stars (``clump giants'') we have
restricted our subgiants planet search primarily to stars with
absolute magnitudes $M_V > 1.8$,
corresponding to masses $M_* \lesssim 2.1$~\msun\
\citep{johnson06a}. Our planet search is therefore complementary to the
survey of \citet{lovis07} and other planet searches around K giants
\citep[e.g.][]{setiawan03, sato03, hatzes05, reffert06, nied07}. The K giants
provide information about massive planets around stars with $M_* \gtrsim
2.1$~\msun, and the relatively stable atmospheres
of subgiants (jitter~$=5$~\ms) allow us to detect planets beyond 1 AU
with minimum masses down to \msini$\approx 0.5$~\mjup.

Our preliminary results and those of \citet{lovis07} reveal
a positive correlation between the mass of stars and the
likelihood that they harbor giant planets. However, testing the exact
shape of the relationship predicted by the simulations of
\citet{kennedy07}---a rising trend of 
planet occurrence up to a peak near 2.5~\msun\ and a decreasing trend
toward higher masses---will require a much  
larger sample of stars. We have expanded our search for planets around 
intermediate--mass stars by adding 300 additional former A and F
stars to our Lick and Keck samples. The results from our expanded planet
search should reduce the error bar on the high--mass bin in
Figure~\ref{mass_hist} by a factor of 2. If the 9\% occurrence
rate for $M_* > 1.3$~\msun\ is confirmed, then our expanded planet
search will result in the detection of 20--30 new planets orbiting
some of the most massive planet host stars in the Solar
neighborhood. In addition to verifying the preliminary results
presented here, these planet detections will also allow us to study
the effects of stellar mass on other planet characteristics such as
eccentricity, semimajor axis and minimum mass. 

\acknowledgements  

We gratefully acknowledge the efforts and dedication of the Lick
Observatory and Keck Observatory staff. We are also grateful to the time
assignment committees of NASA, NOAO and University of California for their
generous allocations of observing time. We thank Ben Zuckerman for his
invigorating conversations regarding M dwarf metallicities. 
We acknowledge support by NSF grants AST-0702821 (to JAJ) and
AST-0307493 (to SSV) and AST-9988087, 
NASA grant NAG5-12182, and travel support from the Carnegie Institution
of Washington (to RPB), NASA grant NAG5-8299 and NSF grant
AST95-20443 (to GWM). DAF is a Cottrell 
Science Scholar of Research Corporation and acknowledges support from
NASA Grant NNG05G164G that made this work possible.
This research has made use of
the Simbad database operated at CDS, Strasbourg France, and the
NASA ADS database.  Finally the authors wish to extend thanks
to those of Hawaiian ancestry on whose sacred mountain of
Mauna Kea we are privileged to be guests.  Without their generous
hospitality, the Keck observations presented herein would not
have been possible.


\begin{deluxetable}{lll}
\tablecaption{Radial Velocities for GJ 317\label{velgl317}}
\tablewidth{0pt}
\tablehead{
\colhead{JD} &
\colhead{RV} &
\colhead{Uncertainty} \\
\colhead{-2440000} &
\colhead{(m~s$^{-1}$)} &
\colhead{(m~s$^{-1}$)} 
}
\startdata
11550.993 &   13.89 & 3.79 \\
11552.990 &   30.27 & 4.99 \\
11582.891 &   45.77 & 4.16 \\
11883.101 &  -23.35 & 3.74 \\
11973.795 &  -65.89 & 6.10 \\
12243.073 &    0.00 & 6.52 \\
12362.949 &   95.56 & 6.07 \\
12601.045 &  -31.64 & 5.09 \\
12989.125 &  101.25 & 4.79 \\
13369.016 &  -29.65 & 3.40 \\
13753.983 &  130.31 & 3.68 \\
14084.001 &   -6.55 & 4.64 \\
14086.141 &  -21.73 & 4.55 \\
14130.082 &  -15.26 & 4.79 \\
14131.014 &  -15.47 & 4.20 \\
14138.932 &   -8.98 & 2.82 \\
14216.733 &    4.15 & 3.86 \\
14255.745 &   28.54 & 2.25 \\
\enddata
\end{deluxetable}

\begin{deluxetable}{lc}
\tablecaption{Stellar Properties for \starA
\label{star_props}}
\tablewidth{0pt}
\tablehead{
\colhead{Parameter} & \colhead{Value} 
}
\startdata
$V$ & \vmagA \\
$B-V$ & \bvA \\
$M_V$ & \mvA \\
$K$  & 7.016 \\
$J-K$ & 0.915 \\
$H-K$ & 0.311 \\
$d$ (pc) & $\dA \pm 1.7$ \\
$M_*$ (\msun) & \mstarA$\pm 0.04$ \\
\fe & \feA $\pm$0.2 
\enddata
\end{deluxetable}

\begin{deluxetable}{lc}
\tablecaption{Orbital Solution for \starA\,b
\label{kep_pars}}
\tablewidth{0pt}
\tablehead{
\colhead{Parameter} & \colhead{Value} 
}
\startdata
$P$ (days) &  \pA $\pm$\peA \\
$P$ (years) &  \pyearsA $\pm$\pyearseA \\
$K$ (m\,s$^{-1}$) & 71.0$\pm$7 \\
$e$ & \eA $\pm$\eeA\\
$T_P$ (Julian Date) & \tpA$\pm$\tpeA \\
Linear Trend (\ms\,year$^{-1}$) & \trendA$\pm$\trendeA  \\
$\omega$ (degrees) & \omA $\pm$\omeA \\
M$\sin i$ (\mjup) & \msiniA \\
$a$ (AU) & \arelA \\ 
N$_{\rm obs}$ & \nobsA \\
RMS (m\,s$^{-1}$) & \rmstr \\
\chisq & \chitr \\
\enddata
\end{deluxetable}
\clearpage

\end{document}